\documentclass[]{spie}  

\usepackage{amsmath,amsfonts,amssymb}
\usepackage{graphicx}
\usepackage[colorlinks=true, allcolors=blue]{hyperref}

\title{Ray-tracing a small orbital mission for soft-X-ray polarimetry}

\author[a]{Hans Moritz G\"unther}
\author[a]{Herman L. Marshall}
\author[a]{Alan Garner}
\affil[a]{MIT Kavli Institute for Astrophysics and Space Research, Massachusetts Institute of Technology, Cambridge, MA 02139, USA}

\authorinfo{Send correspondence to H.M.G. (E-mail: hgunther@mit.edu)}

\begin{document} 
\maketitle

\begin{abstract}
X-ray polarimetry is still largely uncharted territory. With the
upcoming launch of IXPE, we will learn a lot more about X-ray
polarization at energies above 2~keV, but so far no current or
accepted mission provides observational capabilities below 2~keV. We
present ray-tracing results for a small orbital mission that could be
launched within NASA's Pioneer or SmallSat cost-cap to provide X-ray
polarimetry below 2~keV. The design is based on the use of
laterally-graded multi-layer (ML) mirrors, a concept that we have
developed theoretically for the REDSoX Polarimeter\cite{redsox},
for which most components have been verified in the laboratory.  In this
contribution, we describe a single channel orbital mission based on
the same idea, but modified to the unique cost and space
requirements. All results scale up easily to two or more polarimetry
channels. Scaling up would simply increase the effective area and reduce the
need to rotate the instrument to measure the different polarization
directions. In particular, we use the ray-traces to define the maximum
size of the dispersion gratings and to determine an alignment budget.

\end{abstract}


\section{INTRODUCTION}
\label{sec:intro}
X-rays can provide unique insight into the hot and energetic
universe. While X-ray photometry and spectroscopy is now routinely
performed with a number of missions, most prominently XMM-Newton and
Chandra, there is no currently operating observatory to perform X-ray
polarimetry and only for a single source (the Crab nebula) has there
ever been a significant detection of polarized
X-rays\cite{1972ApJ...174L...1N,1978ApJ...220L.117W}.  IXPE\cite{IXPE}
will be launched in 2021 and provide this capability for X-rays above
2~keV, but there is still no instrument in sight for soft X-ray
polarimetry. Yet, a range of open science questions for several
classes of astrophysical sources can only be answered with soft X-ray
polarimetry.

In this paper we present ray-trace studies for a a small orbital mission for soft x-ray polarimetry, which is described in section~\ref{sect:mission}. In section~\ref{sect:raytrace} we describe the setup of our ray-trace simulations. From those, we can predict the performance, most importantly effective area and modulation factor (section~\ref{sect:performance}, perform trade-studies on shape and the size of the gratings used in the design (section~\ref{sect:trades}, and evaluate the influence of alignment errors and other non-ideal effects (section~\ref{sect:align}). We end with a short summary in section~\ref{sect:summary}.

\section{MISSION OVERVIEW}
\label{sect:mission}
Over the years, we have developed and refined a design for a soft
X-ray polarimeter, called ``REDSoX'', which is based on reflection off
a laterally-graded multi-layer (ML) mirror.  Details of the REDSoX
design are discussed in Ref.~\citenum{redsox}, more details of
ray-trace calculations are presented in Ref.~\citenum{redsoxtrace}.
In short, X-rays are focused into a converging beam using a
mirror. Beyond the mirror, the photons encounter critical-angle
transmission (CAT)
gratings\cite{Heilmann:11,doi:10.1117/12.2188525,10.1117/12.2314180,10.1117/12.2529354}. A
fraction of the photons passes straight through the gratings onto an
imaging detector, which can be used to confirm the accuracy of the
pointing and provide an X-ray spectrum with the intrinsic energy
resolution of the detector. Most photons, however, are diffracted. By
selecting the blaze angle (the angle between the grating normal and
the direction of the incoming photons), we can optimize the fraction
of photons diffracted into the first order. Photons of longer
wavelengths are diffracted farther and thus hit the focal plane farther
away from the direct image than photons with shorter wavelengths.
A multi-layer mirror is located in
the focal plane. The thickness of the
layers varies with position (``laterally graded'') and is chosen such
that every photon interacts with the mirror at a position where the
mirror spacing matches Bragg condition for photons of that
wavelength\cite{redsox,redsoxtrace}. The mirror is tilted by about 45
degrees with respect to the incoming photons, and thus only photons
with one polarization direction are reflected (``Brewster angle''),
while photons with a perpendicular polarization are absorbed. A CCD
detects the reflected photons. The requirement to match the photon
position to the Bragg peak on the multi-layer mirror sets a limit to
the width of the mirror point-spread function in the dispersion
direction. We achieve this by sub-aperturing, i.e.\ we use only
wedge-shaped mirror modules and not full circles.

Measurements at several angles are required to reconstruct the
polarization fraction and angle. In our design, this can be achieved
with a single multi-layer mirror if the entire instrument rotates
(either continuously or in steps) throughout the observation or with
two or three channels, each with its own sub-aperture, multi-layer
mirror and detector, that are rotated with respect to each other.

REDSoX is designed as a sounding rocket payload with very short
exposure times, but relaxed mass and volume requirements. In this paper,
we present ray-traces for a variant of the REDSoX design optimized for
a small orbital mission, e.g.\ in NASA's Astrophysics Pioneer or
SmallSat program. Compared to REDSoX, we need to shorten the focal
length to fit the mission into the envelope of a secondary payload,
such as when mounted to an ESPA grande ring.
While the mirror area may be smaller, we can integrate for
days to months in an orbital
mission while a sounding rocket gives only a few minutes of observing
time.

Most of the trades and studies done for REDSoX\cite{redsoxtrace} are
applicable here, too, just with different numbers for the focal
length, but others are unique to this design, e.g.\ the much longer
exposure times require a much more careful estimate of the background
that is picked up in the detectors.  While the sounding rocket REDSoX
is designed with three polarization channels, an orbital mission with
much longer exposure times can work with just one or two channels
because lower effective areas can be compensated by longer exposure
times.

\begin{figure} [ht]
  \begin{center}
    \includegraphics[height=5cm]{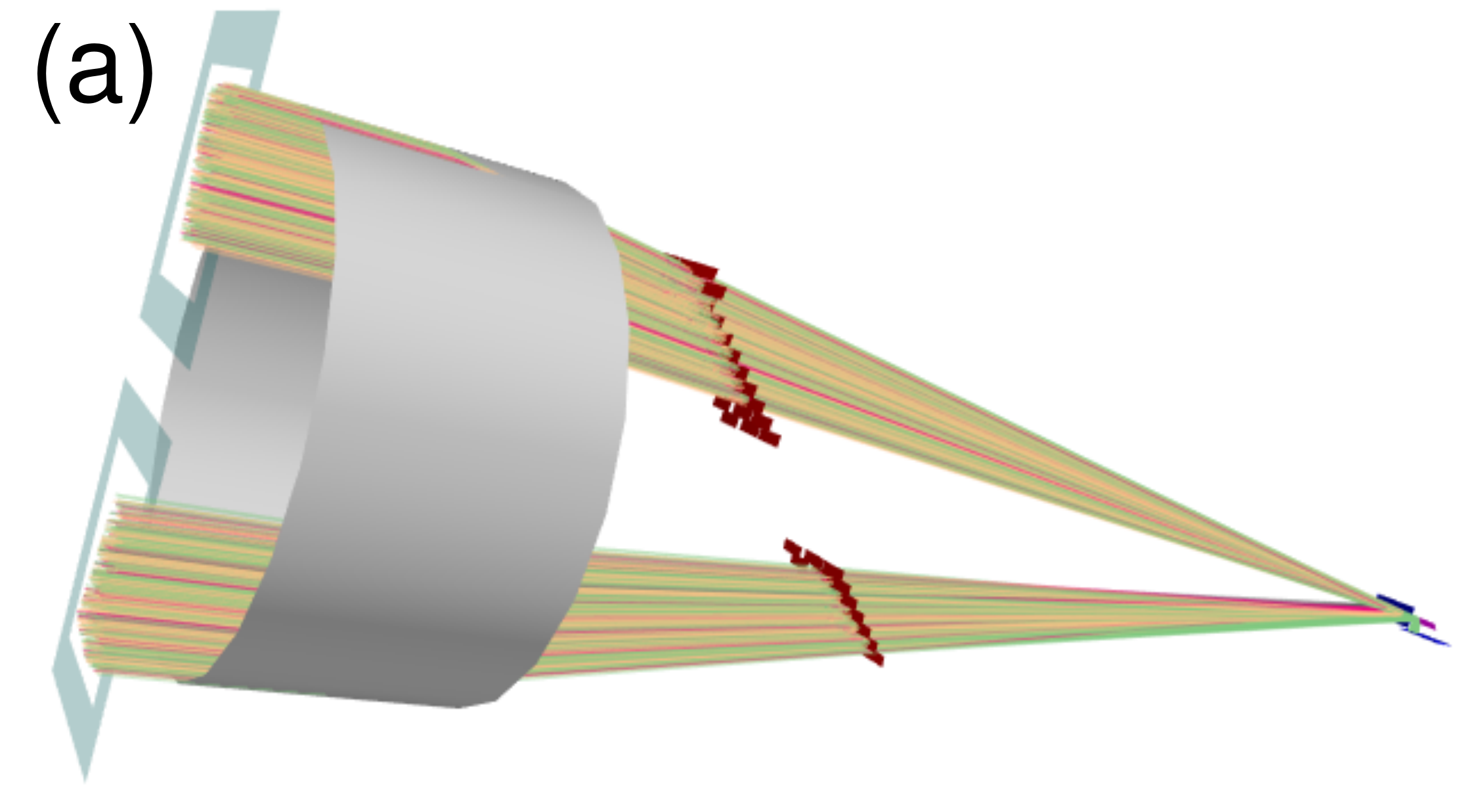}
    \includegraphics[height=5cm]{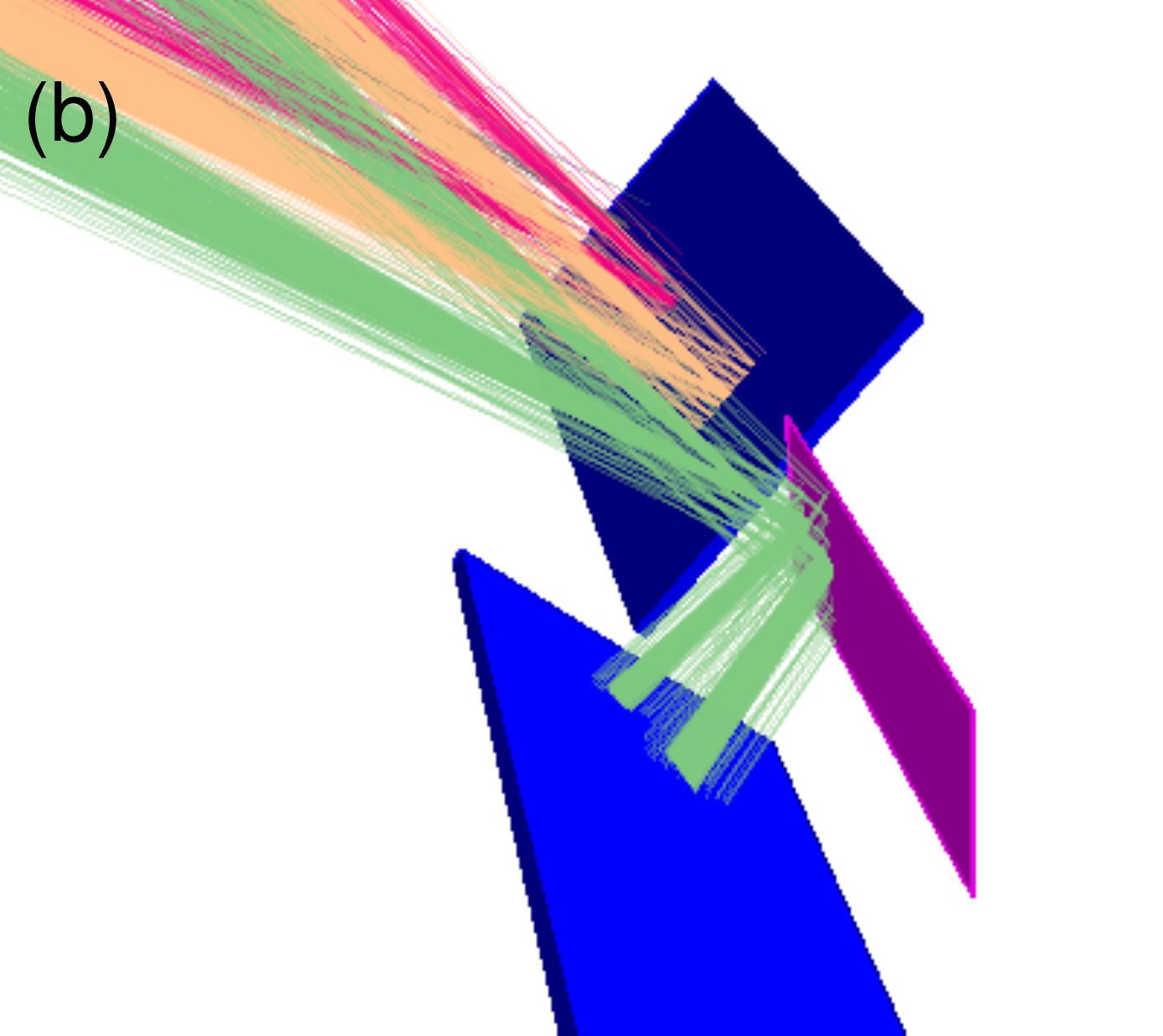}
  \end{center}
  \caption
      { \label{fig:3d}3D rendering of the instrument design and a
        ray-trace for monoenergetic rays of 0.3~keV. The ray-trace
        setup makes some simplifications. In particular, the mirror is
        not modeled in 3D, but approximated by a 2D lens. The position
        of the mirror modules is indicated by a cylinder that has the
        same radius as the outermost mirror surface. This is a
        monochromatic simulation with photon energies of 0.3 keV. Rays
        are colored according to the grating diffraction order. The
        zeroth order is shown with orange rays, the first order with
        green rays. The green rays bounce of the multilayer-mirror
        (purple) before they hit a detector (blue). The second
        detector (also blue) images the zeroth order, but also photons
        diffracted into order -1 (red rays). \emph{panel a:} View of
        the full instrument. The astrophysical source is located
        towards the left. Rays are shown starting at the entrance
        aperture. \emph{panel b:} Zoom into the focal plane with
        multi-layer mirror (purple) and two detectors (blue).  }
\end{figure}

In this work, we consider a system with a focal length of just 1.25~m
and a single polarimetry channel. An overview of the design is shown
in Figure~\ref{fig:3d}. This matches our submitted Astrophysics
Pioneer proposal (dubbed ``PiSoX'' for Pioneer Soft X-ray
polarimeter) and represents the simplest possible instrument to
measure soft X-ray polarization. It is easy to extend this design by
adding another polarimetry channel for redundancy, increased effective
area, and the opportunity to measure fast changes in the polarization
(since two polarization directions are observed simultaneously,
instead of rotating the instrument), but for simplicity we will call
the instrument ``PiSoX'' for the remainder of this article.

Figure~\ref{fig:3d} shows a ray-trace through our one-channel
system. All rays in the figure have the same energy (0.3~keV). Because
the mirrors deliver a converging beam, rays have different angles with
respect to the ML mirror. Because of that, they require a slightly
different period of the layers of the ML mirror to match the Bragg
peak, and consequently, they have to be diffracted to a different
position on the ML mirror. This is the origin of the stair-stepped
arrangement of the CAT gratings\cite{redsox,redsoxtrace}. In panel
(b) of the figure, it can be seen that the rays fall into two groups
depending on if they pass through the upper or lower sector of
gratings.

\section{SETUP FOR RAY-TRACES}
\label{sect:raytrace}
We perform a geometric ray-trace with the MARXS code, which follows
individual rays through the system from the entrance aperture to the
detector. Simulations are performed with MARXS
1.2\cite{marxs,marxs1.2}. MARXS is written in Python and available
under the GNU license v3. The code is available on
github\footnote{\url{https://github.com/chandra-marx/marxs}}. Specific
code for the simulations of the instrument shown here is also
available\footnote{\url{https://github.com/X-raypol/ray-trace}}; we
used the version with commit hash 13b10be. More details on the code
can be found in proceedings describing its application for other
missions\cite{10.1117/12.2525814,10.1117/12.2312678}.

Every simulation contains some simplification of reality, e.g.,
because certain aspects of the design are not known yet, laboratory
data about performance must be extrapolated, or simply because
computational cost puts limits on how many photons can be run through
a simulation. Our simulations are set up with a relatively simplistic
mirror model. Instead of a three dimensional structure, the mirror is
implemented as plane perpendicular to the optical axis. When rays
intersect the plane, their directions are modified as if they pass a
perfect lens. In the next step, additional scatter in the plane of
reflection is added. For each ray, the scattering angle is drawn from
a Gaussian distribution, where the width of the distribution is tuned
to achieve a PSF size that matches the PSFs measured or expected for
the type of mirror used. We use 12 arcsec for the $\sigma$ of the
Gaussian.

Our instrument uses CAT gratings manufactured from Si wafers at the
MIT Space Nanotechnology Laboratory
\cite{Heilmann:11,doi:10.1117/12.2188525,10.1117/12.2314180,10.1117/12.2529354}. The
grating efficiency of these gratings is calculated based on
simulations and verified in the laboratory; a table of efficiencies is
an input to our simulations. The high aspect-ratio grating bars are
4~$\mu$m deep and supported by an L1 support structure running
perpendicular to the grating bars themselves and the entire membrane
(bars and L1) is mechanically stabilized by a hexagonal L2 support
structure, which is 0.5~mm deep. Absorption and diffraction of photons
by the L1 and L2 support structures is included in our
simulations. The grating membrane is surrounded by a 1.5~mm wide frame
of solid Si which is glued into a grating holder. We discuss in
section~\ref{sect:bend} that we require the $30\times10$~mm sized
gratings to be bend along the long axis. Laboratory measurement show
that bending does not decrease performance\cite{10.1117/12.2274205}
and grating alignment has also been
proto-typed\cite{10.1117/12.2314902,10.1117/12.2529607}.

Laterally graded multi-layer (ML) mirrors can be made from a variety
of materials. We use a Cr/Sc mirror, because this combination of
material promises a good reflectivity in the bandpass of interest; ML
mirror reflectivities have been verified in the
laboratory\cite{10.1117/12.2188452}. The detectors will be CCD ID-94
detectors produced by Lincoln Labs, we use the quantum efficiencies
measured for Suzaku CCDs\cite{2007PASJ...59S..23K} for our ray-traces.

\section{PERFORMANCE}
\label{sect:performance}
The main characteristics of a polarimeter are its effective area and the modulation factor.

\subsection{Effective area and modulation factor}
\begin{figure} [ht]
  \begin{center}
    \includegraphics[height=5cm]{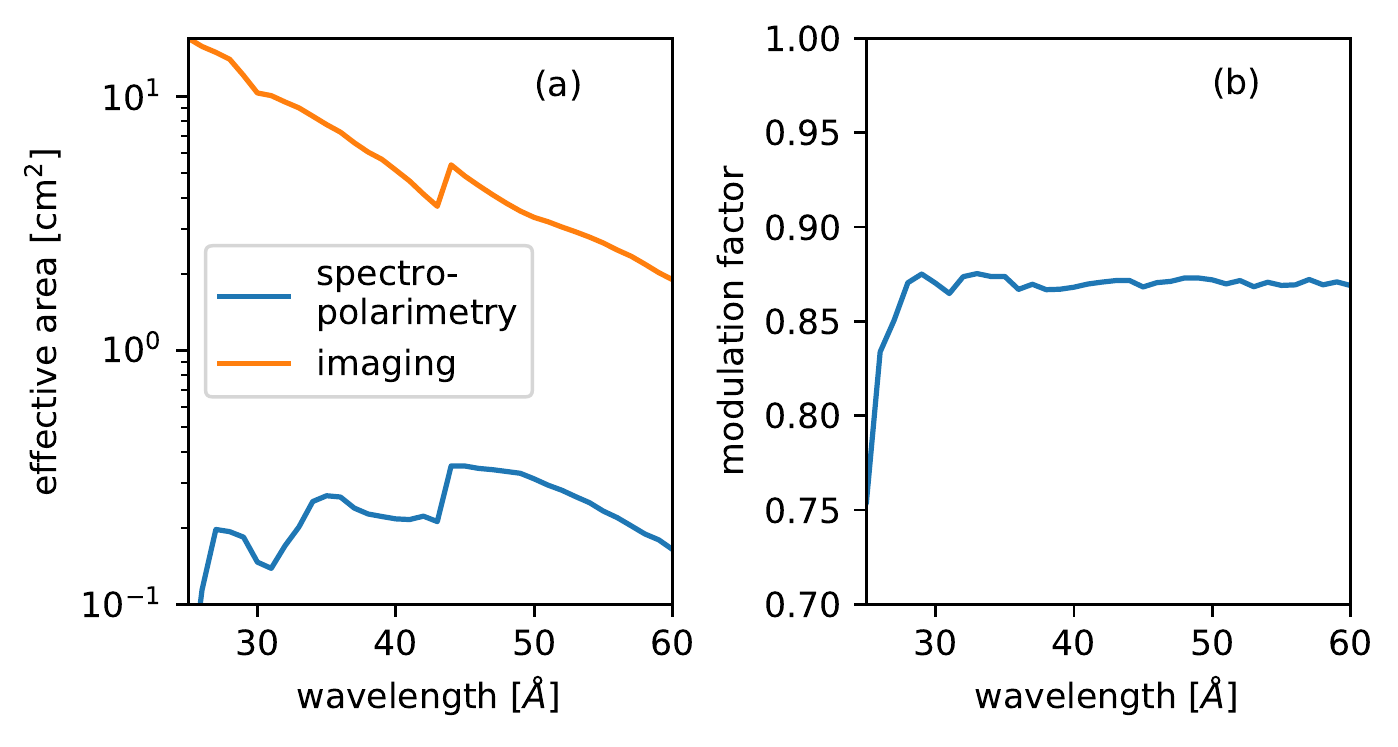}
  \end{center}
  \caption
      { \label{fig:aeff} \emph{panel a}: Effective area. CCD~1 detects
        the polarized signal, CCD~0 the zeroth order. Because CCD~1
        observes rays reflected off the ML mirror, the effective area
        is much lower than in the CCD~0. \emph{panel b:}: modulation
        factor.  }
\end{figure}

In order to determine the modulation factor, we perform simulations
for a source that is 100\% polarized. Then, we rotate the pointing of
the instrument on the sky in small steps so that we can build up the
modulation curve. From this curve, the total modulation of the signal
can be calculated as $\frac{T-B}{T + B}$, where $T$ is the maximum of
the curve and $B$ is the minimum. The energy dependence for effective
area and modulation are shown in figure~\ref{fig:aeff}.

\subsection{How does an observation look?}
Since PiSoX has only a single polarimetry channel, we have to rotate
the instrument on the sky to cover enough range in rotation angles to
sample all possible polarization directions. We show simulations with
the spectrum and flux of Mk 421 assuming the source is fully
polarized. In reality, the polarization fraction is going to be $<1$
and might depend on energy. Of course, we can simulate those
scenarios, too, but the point here is to show how the data works in
principle, so we pick the easiest scenario.

We simulate two observations with different polarization angles on the
sky, which are chosen such that the first angle gives the maximal
signal and the second one the minimal signal on the polarization
channel. In a real observation, that angle is not known {\it a priori} and
the instrument needs to be rotated continuously or observe at at least
three angles to be able to uniquely derive the polarization fraction
and angle from the polarization channels alone. Using the zeroth order
data as well, then observations at two angles are sufficient.

\subsection{Background}
For faint sources, background limits the signal-to-noise ratio that
can be achieved with very long observations. There are two main
components to the background: Particle background and astrophysical
sources. We assume that the particle background is
independent of the orientation of
the spacecraft.
With CCD detectors, the main mechanism for reducing background
is to select data using the
shapes and energies of the detected events. For a
dispersive spectro-polarimeter like
PiSoX only background events that match the energy of the photons
expected at that position on the detector are relevant, reducing the
influence of the particle background. Astrophysical background stems
from the Cosmic X-ray Background, composed of a soft, thermal Galactic
foreground component and power-law component to account for unresolved
AGN. The spectral shape and flux of the astrophysical
background are taken from an analysis of Abell
1795\cite{2009PASJ...61.1117B}.  We simulate this as a disk of
1~degree radius (limited by the field-of-view of the thermal
pre-collimators) to determine how diffuse sky background influences
the count rate in the polarization channel. Since
dispersed source count rates can be very low, background can
potentially become important. On the other hand, the multi-layer
mirror will only reflect photons coming with a very specific
combination of angle and energy, reducing the background
significantly.

For a 100 ks observation, we predict 0.2~cts in the polarimetry
channel when extracting the chip area that contains the signal of an
on-axis point source. This is an upper limit, because it does not
yet consider energy filtering using the intrinsic CCD resolution, which
may reduce the background rate by another factor of two or so. The
simulation predicts of order 100 counts that spatially overlap with
the zeroth order of an on-axis point source, where the exact number
scales with the size of the extraction region. The difference in flux
between imaging and polarimetry channel is $10^4$, while the effective
area differs only by one order of magnitude. From this, we can
conclude that the ML mirror suppresses the background by about 
a factor of $10^3$.
At a flux of $2\frac{\mathrm{ct}}{\mathrm{Ms}}$, we can safely assume that
diffuse X-ray emissions is negligible, as several thousand
counts are needed to detect polarizations below 10\%.

\section{TRADES}
\label{sect:trades}

\subsection{Bending gratings}
\label{sect:bend}

\begin{figure} [ht]
\begin{center}
\includegraphics[width=0.5\textwidth]{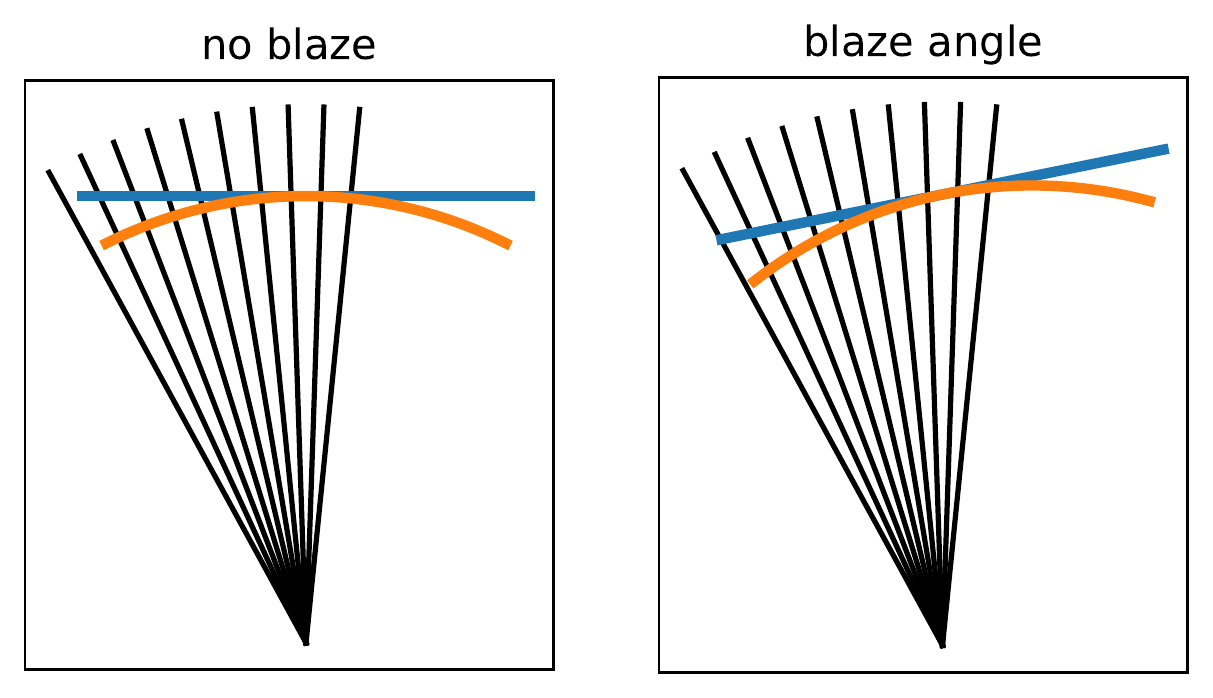}
\end{center}
\caption {\label{fig:explainbending}
In a converging beam (black rays) the angle between the grating normal and the rays changes with grating position for a flat grating (blue). Bending the grating can compensate for this problem (orange) for gratings that are small compared to the radius. This is true for gratings where the ray through the grating center is normal to the grating (left) and for gratings with a design blaze angle (right).
}
\end{figure}

\begin{figure} [ht]
   \begin{center}
   \includegraphics[height=5cm]{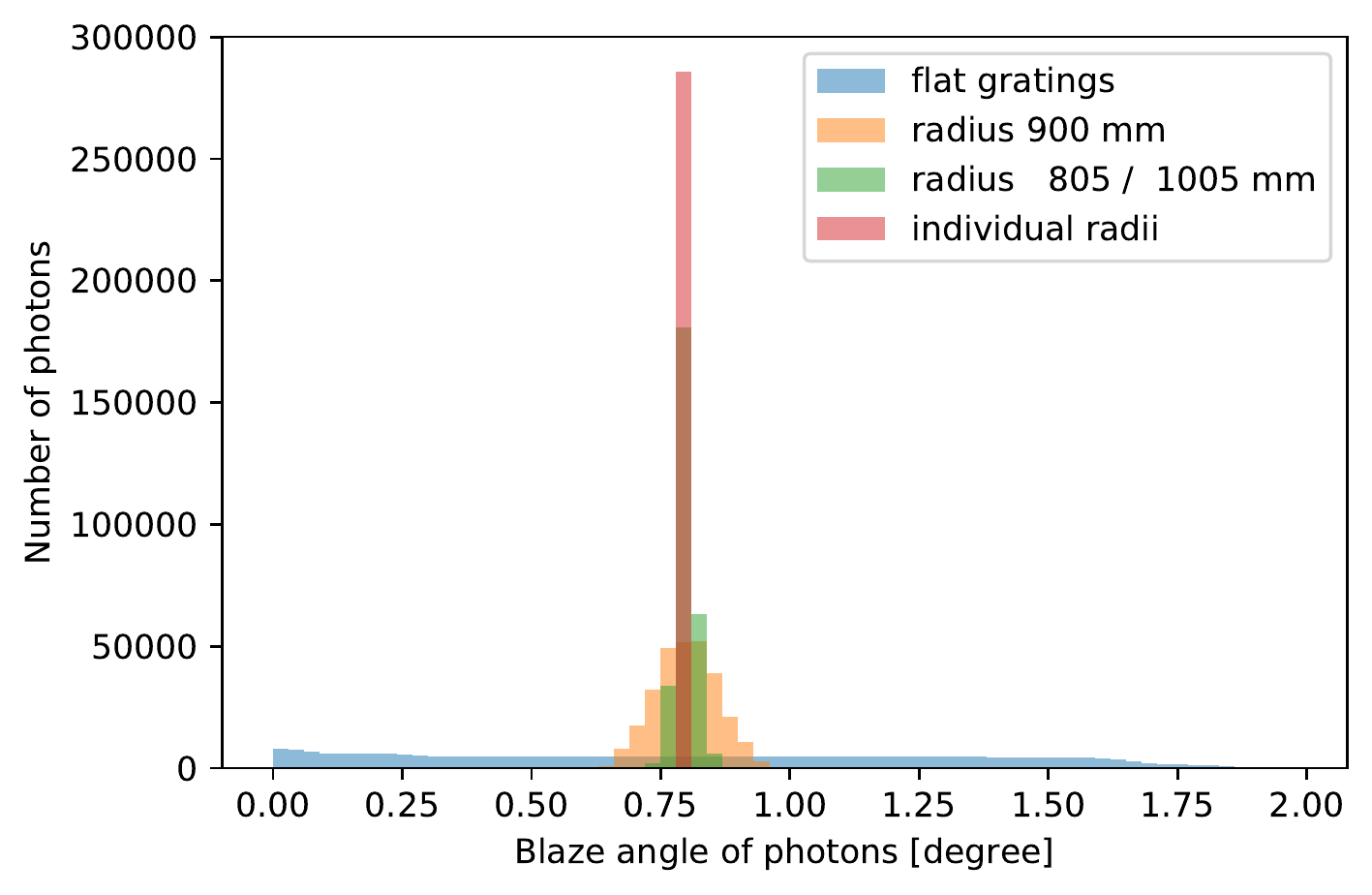}
   \end{center}
   \caption
   { \label{fig:curvature}Distribution of blaze angles (angle between incoming ray and local normal to the grating surface) for different radii of bent CAT gratings.
}
   \end{figure}

The grating efficiency of CAT gratings changes with the blaze
angle. If the blaze angle is small, i.e.\ the rays hit the grating
parallel to the surface normal, then many rays will end up in the
zeroth order. Higher blaze angles favour higher orders. PiSox is
designed for a blaze angle of 0.8~degree to maximise the number of
photons that are diffracted into the first order, where they will hit
the multi-layer mirror at the Bragg peak. We position the gratings
such that a ray hitting the grating center has the correct blaze
angle. However, the gratings are located in a converging beam and thus the blaze
angles for rays hitting a flat grating near its edges are different from
the nominal blaze angle (Figure~\ref{fig:explainbending}. Fewer photons are diffracted into the first
order and the effective area is reduced.
If the grating surface
follows a cylinder with a radius of curvature that matches the
distance of the grating from the focal point, that effect can be
almost entirely be compensated (Figure~\ref{fig:curvature}). The axis
of that cylinder is parallel to the cross-dispersion direction. In
other words, the grating is curved along the long side. On the other
hand, bending almost every grating with a different radius of
curvature increases the complexity and thus cost and schedule risk
dramatically. Here, we study four different scenarios:
\begin{itemize}
    \item Gratings are flat.
    \item All gratings are bent with the same radius which is chosen to be close to the average distance between the gratings and the focal point.
    \item We use two different radii of curvature for the upper and lower part of the grating staircase.
    \item Each grating is curved individually.
\end{itemize}

\begin{table}
  \caption{Effective area in the imaging and polarimatery channel for a representative energy of 0.277 keV}
  \label{tab:curvature}
\begin{center}
\begin{tabular}{c|c|c}
scenario & imaging & polarimetry \\
& $\mathrm{cm^{2}}$ & $\mathrm{cm^{2}}$ \\
\hline\hline
flat gratings & 5.43 & 0.145 \\
radius 900 mm & 4.99 & 0.210 \\
radius   805 /  1005 mm & 4.99 & 0.214 \\
individual radii & 4.97 & 0.213 \\
\end{tabular}
\end{center}
\end{table}
Table~\ref{tab:curvature} shows simulated effective areas at a
representative energy for all four scenarios.  Flat gratings have the
best imaging performance, but severely reduce the effective area of
the more important polarimetry channel. There is little performance
difference between the different bending options studied here, so we
conclude that the simplest option (all gratings have the same
curvature) should be the baseline design for our instrument.

\subsection{Grating dimensions}
We use a baseline design with rectangular CAT gratings with edge
lengths of 30~mm and 10~mm, where the long edge is parallel to the
dispersion direction. In this trade, we test the choice of 10~mm width
for the cross-dispersion direction. CAT gratings can be manufactured
in larger sizes, but the ideal surface on which the gratings need to
be placed is saddle-shaped. On the other hand, the design of the
gratings requires the grating normal to be roughly perpendicular to
the incoming rays. If the angle between ray and grating becomes large,
then the support structures that hold the grating bars in place, in
particular the L2 support,
would cast large shadows and thus reduce the effective area. The
grating membranes are fixed to the metal grating holder, which in turn
is mounted to a larger mechanical structure. All these block some
fraction of the area and thus reduce the number of rays that make it
through to the detector.

So, there are two competing effects: On the one hand, larger gratings
reduce the number of and thus area lost to these mounting structures;
on the other hand, CAT grating
normals have to be close to the direction of the incoming rays. Thus, the
larger the grating, the more the regions on the edges deviate from
the surface on which the diffraction should happen. Instead, some rays
hit the grating where it is located ``above'' the theoretical surface,
and some ``below''. In the the first case, photons are diffracted too
far, in the latter too little. Both can cause a photon to arrive
at a position on the
ML mirror that does not match the Bragg peak for its
wavelength.

\begin{figure} [ht]
  \begin{center}
    \includegraphics[height=5cm]{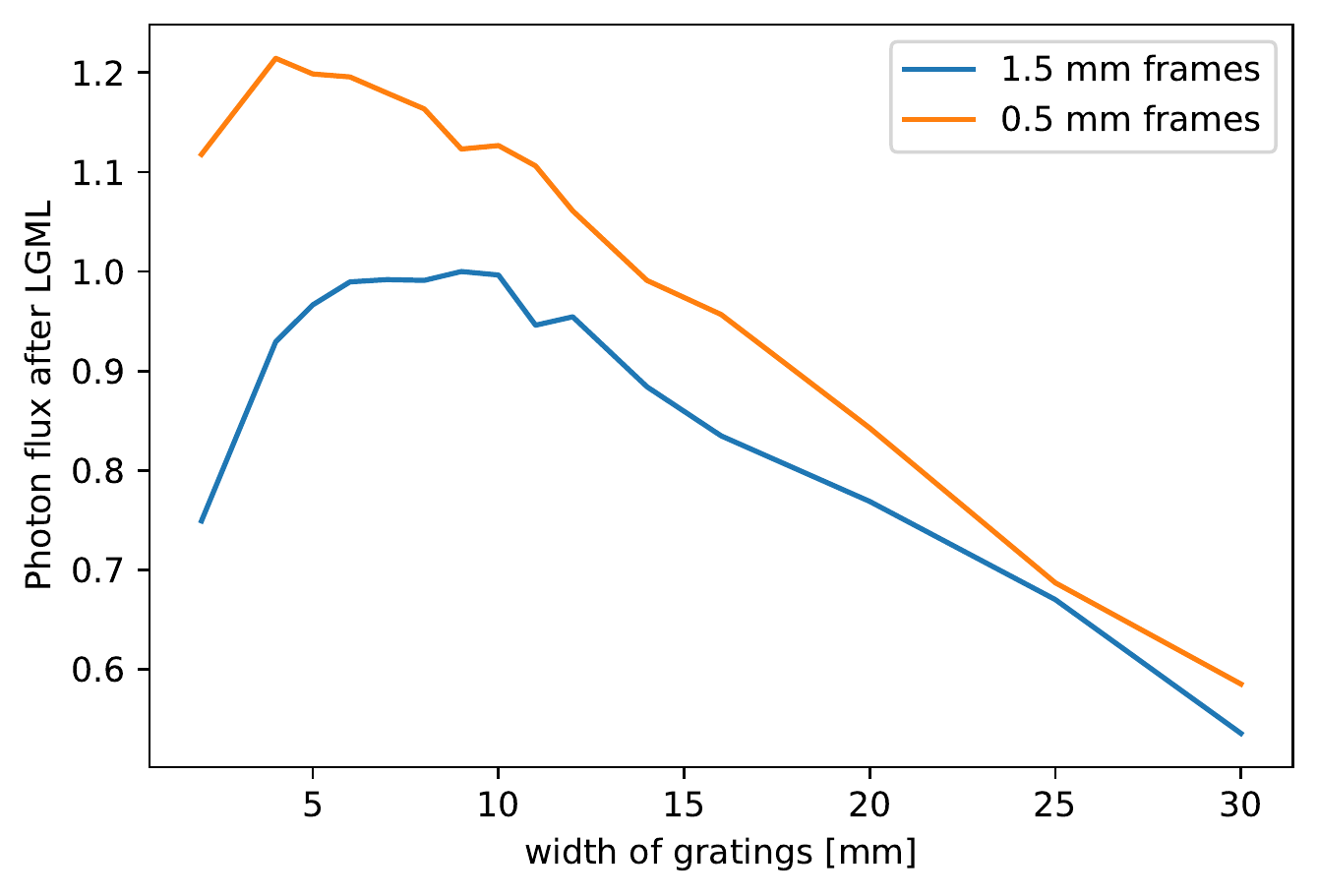}
  \end{center}
  \caption
      { \label{fig:width}Effective area for an instrument filled with
        CAT gratings of different dimensions. In all simulations, the
        CAT gratings are 30 mm long in dispersion direction, but the
        width of the grating in cross-dispersion direction
        differs. When using larger gratings, there is less obscuration
        by frames and grating holders, but gratings deviate more from
        the ideal surface. The figure shows that gratings with
        dimensions of $30\times10\;\mathrm{mm}^2$ perform best.  }
\end{figure}

Figure~\ref{fig:width} shows the effective area in simulations with
different CAT grating sizes. For grating holders with wide frames the
ideal width is between 5 and 10~mm. Since the total effective area
plateaus in that range, we can see that 10~mm wide gratings are a good
choice for the design baseline. Narrower gratings can achieve the same
effective area, but at a larger cost and complexity.

\section{ALIGNMENT AND ERROR BUDGET}
\label{sect:align}
When the physical hardware for a mission is put together, nothing is
perfect. Parts and pieces will always differ slightly in form, shape,
and position from the locations assigned to them in the abstract
design model. Ray-tracing is one useful method to study how much such
misalignments will impact the performance of the instrument and thus
to develop a table of alignment requirements. The looser the
requirement can be, the cheaper and faster the process is. On the
other hand, if the ray-tracing shows that certain elements need to be
positioned very precisely, specific alignment procedures and tests
might have to be developed.

For each alignment parameter, we study six degrees of freedom (three
translations and three rotations). In practice, misalignments happen
in all six degrees of freedom for all parts of the instrument at the
same time. However, computational limitations prohibit us from
exhaustively exploring the full parameter space. Instead, as a first
phase, we treat PiSoX as a hierarchical collection of many elements
(mirror shells, mirror module, CAT gratings, CAT grating assembly, all
of which combine into the optics module etc.). We perform simulations
for about a dozen elements and for each parameter we typically run
simulations for 10-20 values. The full parameter space would thus
require $20^{6*12}=5\times 10^{93}$ simulations. Instead, as a first
step, we set up a perfectly aligned instrument and then vary one
parameter for one element or one group of elements (e.g. the $x$
position of all gratings) at a time. Note that even a perfectly
aligned instrument has some limitations that are inherent in the
design, such as optical aberrations. We step through different values
for each parameter, keeping all other alignments perfect, and run a
simulations with 100,000 photons for each step. We inspect the results
from simulations and select a value for the acceptable misalignment in
each degree of freedom, e.g.\ the value where the effective area of the
channel degrades by no more than 10\%. Selecting the exact value is a
trade-off with engineering concerns. In some degrees of freedom, the
alignment may be easily reached by machining tolerances and thus we
can chose a number that causes only a negligible degradation of
performance, while in other cases, reaching a certain alignment might
be very costly and thus we want to set the requirements for these
parameters as loosely as possible.

The effects of different misalignments will add up in some cases. In others they may cancel each other out to some
degree or combine multiplicatively. In future work, we will
investigate misalignments for all parameters simultaneously.
For the purpose of developing the error budget, there are other design
parameters that are not technically related to mechanical alignment, but
impact the performance in a similar way and can be analyzed with the
same ray-trace setup. One example is the pointing jitter, which
describes how uncertainties in the instrument pointing on the sky
degrade the instrument performance. If the pointing direction on the
sky jitters with time, photons will not always arrive on-axis. This is
somewhat similar to a misaligned optics module.

In this section, we run simulations varying one degree of freedom at a
time. Parameters that are not mechanical misalignments, such as the
pointing jitter, are shown in plots with a blue background in the
coming figures. Simulations for mechanical misalignments come in two
flavors. Either an entire set of objects is moved deterministicly
(e.g. all gratings in the grating assembly are moved 1 mm to the
right) or a number of objects are moved randomly (e.g.\ all gratings in
the grating assembly are moved along the $x$-axis, but for each
grating a new number is drawn from a Gaussian distribution with
$\sigma=1$ mm). The first case is shown with a gray background, the
latter case is shown with a light red background. Results for all mechanical
tolerancing are shown as sets of six plots. The upper row presents
results from translations along the $x$, $y$, and $z$-axis, the bottom
row rotations. The center of the rotation is typically the center of
an element. The coordinate system for the instrument places the optical axis
along the $z$-axis with photons coming in from $z=+\infty$. The origin
of the coordinate system is at the nominal focal point of the mirror
system. The dispersion direction of the gratings is along the positive
$y$-axis. Thus, the long axis of the ML mirror is also parallel to the
$y$-axis.

\subsection{Pointing and mirror}
\begin{figure} [ht]
  \begin{center}
    \includegraphics[height=5cm]{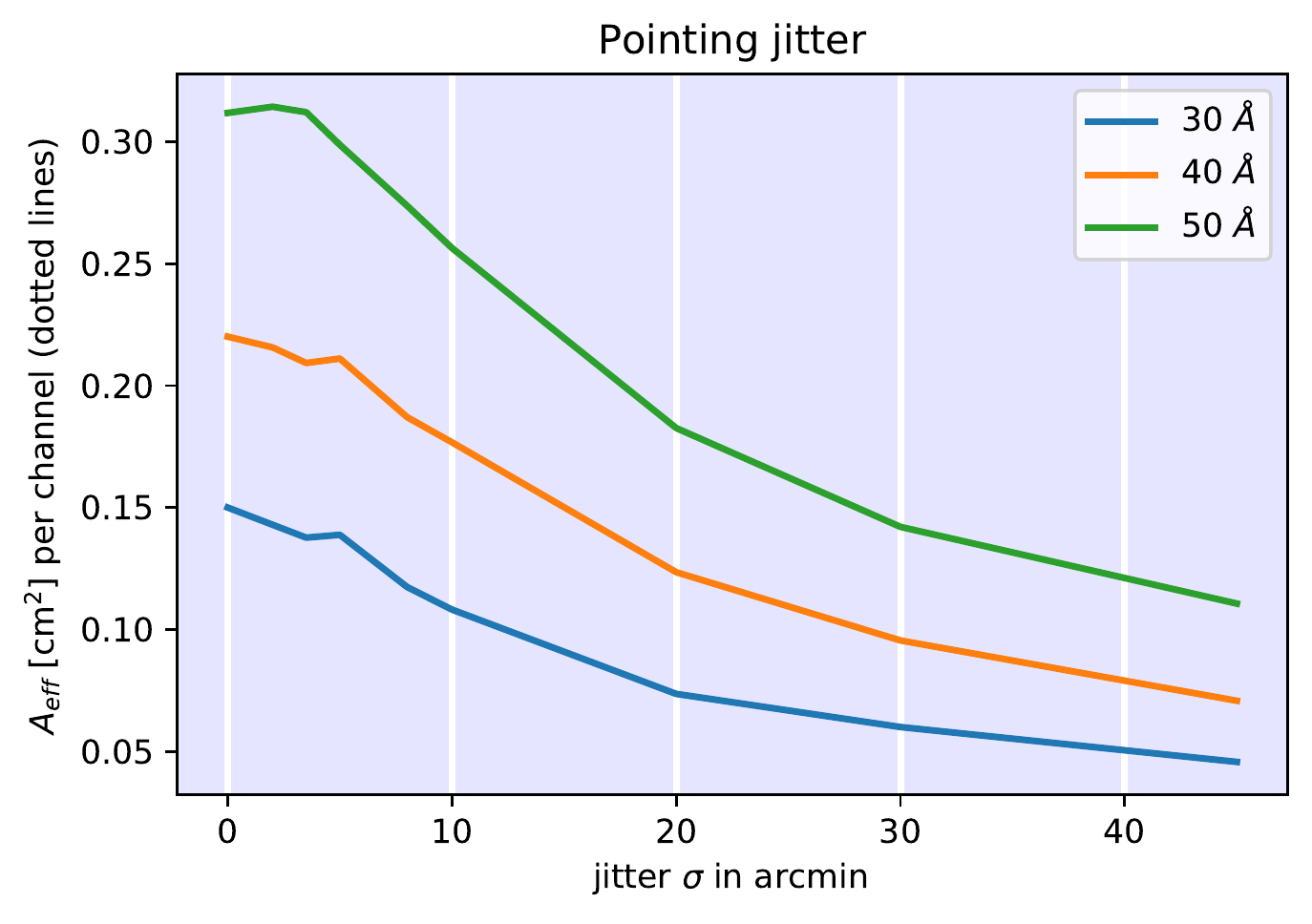}
  \end{center}
  \caption
      { \label{fig:jitter}Change of effective area with increasing pointing jitter. 
}
\end{figure}

Figure~\ref{fig:jitter} shows simulations using an unsteady
pointing. The average pointing direction is on-axis, but the pointing
jitters around that. For each photon, the true pointing direction is
drawn from a Gaussian with the $\sigma$ given in the figure. This
jitter represents uncertainty in the pointing, which can come from
different sources, such as limited resolution of the star trackers,
motion of the pointing within the time period of reading out the star
trackers or integration time of the zero-order image (if used to
determine the target coordinates), or the spacecraft not correcting a
pointing drift fast enough.

\begin{figure} [ht]
  \begin{center}
    \includegraphics[height=5cm]{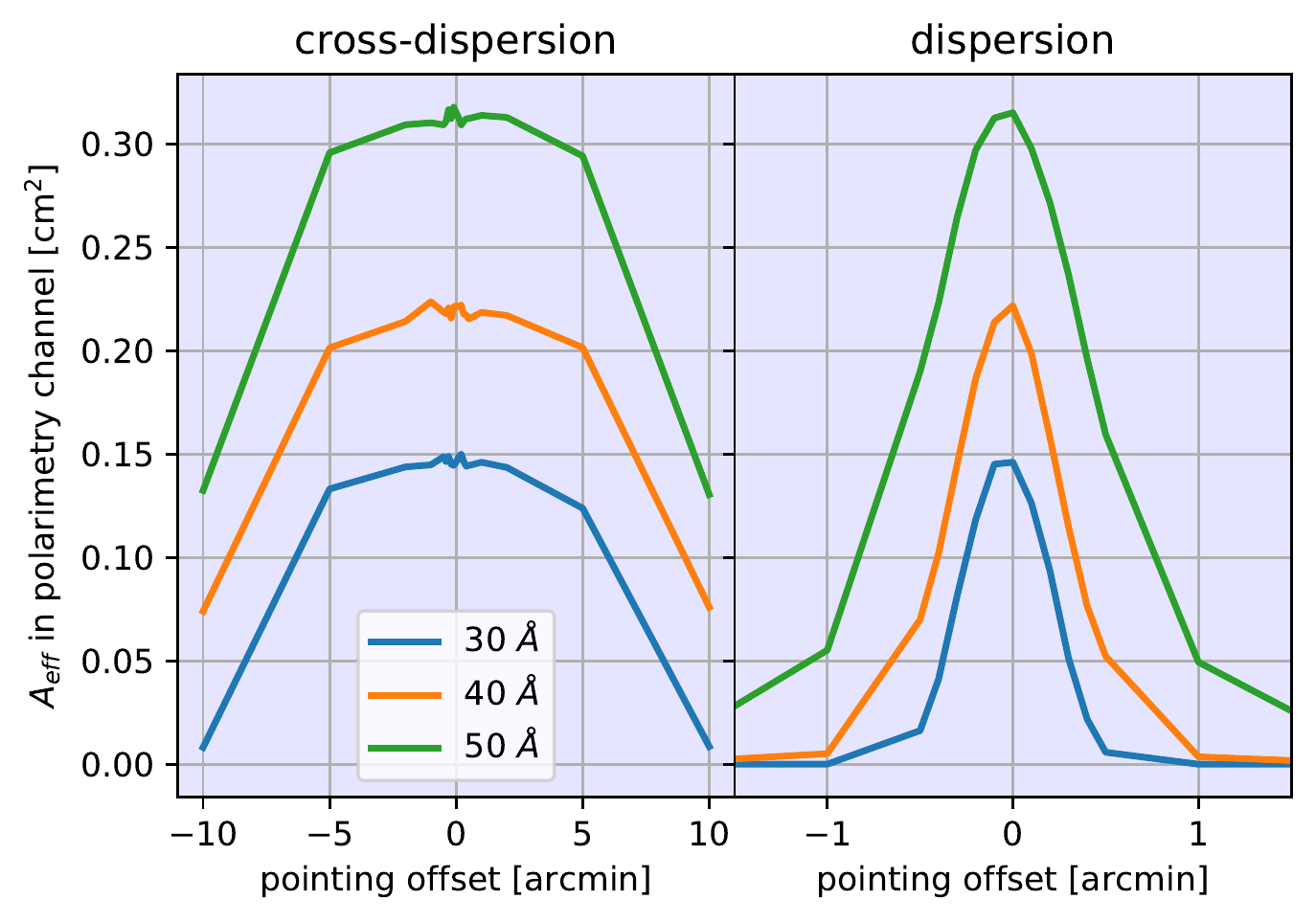}
  \end{center}
  \caption
      { \label{fig:offset_point}Change of effective area for observations where the target is positioned offset from the nominal pointing direction. 
}
\end{figure}

The effective area $A_{\mathrm{eff}}$ drops with increasing jitter,
because the diffracted photons do not hit the ML at the position of
the Bragg peak when the target is not at a nominal position, and thus
the reflectivity is lower. The drop becomes important for a jitter
above a few arcminutes. Mispointing along the direction of diffraction
has a much stronger effect than perpendicular to it. This is
investigated in figure~\ref{fig:offset_point}.

The modulation factor changes only marginally when the source is
observed offset from the nominal position. However, as explained for
the simulations with the pointing jitter above, the effective area
drops dramatically when the source moves along the axis of the ML,
because that means that photons will no longer arrive at the position
where the spacing of the ML matches the required number given the
angle and wavelength of the photon.

\subsection{CAT gratings}
\begin{figure} [ht]
  \begin{center}
    \includegraphics[height=8cm]{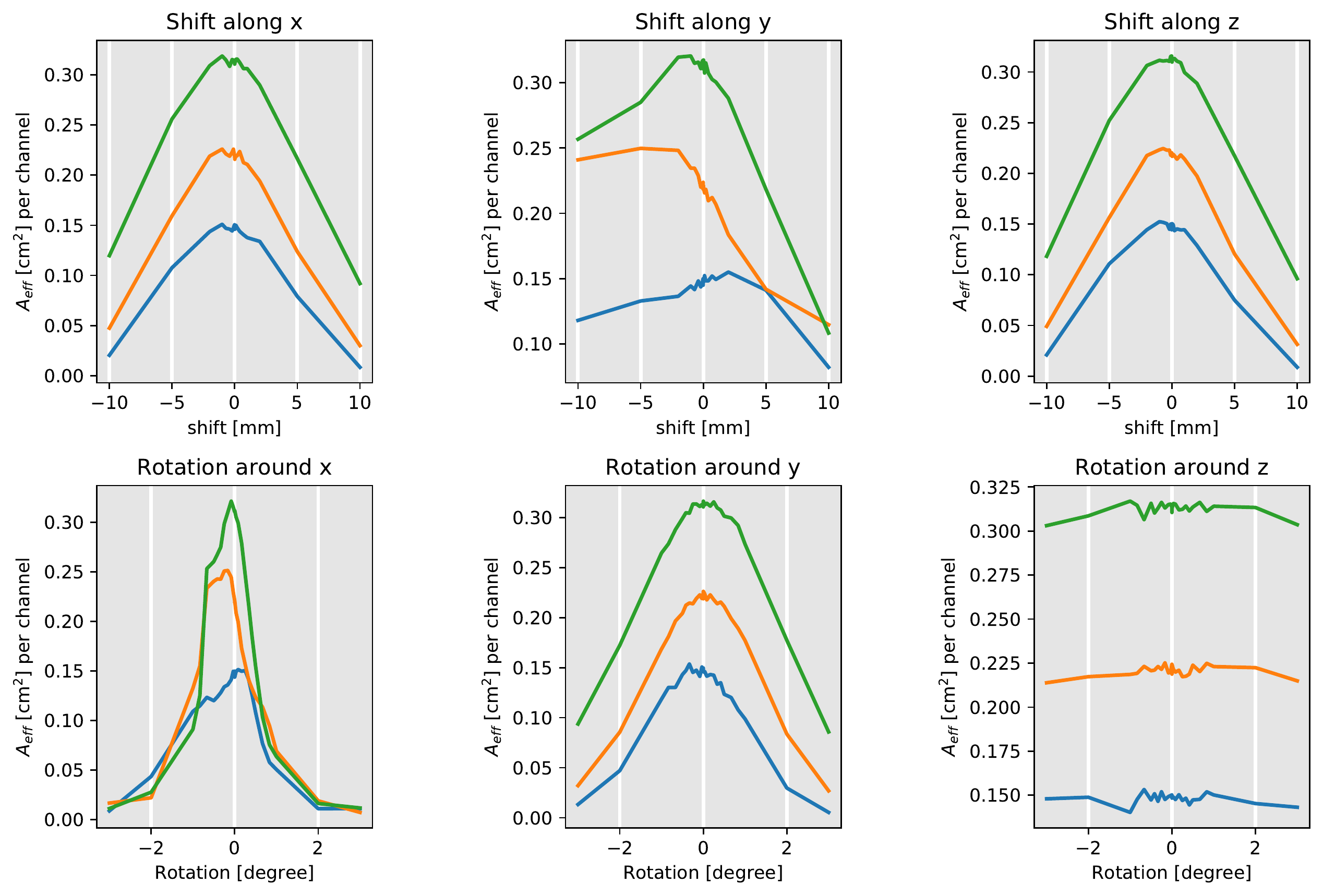}
  \end{center}
  \caption
      { \label{fig:CAT_global}Change of effective area for globally misaligned CAT grating module. 
}
\end{figure}
Figure~\ref{fig:CAT_global} shows simulations that move the CAT
grating module as a whole, i.e.\ translation in $z$ means that all
gratings of both sectors are moved up or down together. This
particular case changes the distance between the gratings and the
focal plane and thus photons will hit the ML mirror on a different
location. Changes of more than a few mm will cause the photons to miss
the position of the Bragg peak on the ML mirror and thus reduce
$A_{\mathrm{eff}}$. The layout is insensitive to translations in $y$
(along the dispersion direction). This is the long direction of the
CAT gratings and CAT gratings are tilted only ever so slightly, so
that the point of intersection is essentially
constant. $A_{\mathrm{eff}}$ only begins to drop when gratings
moved so far that some fraction of the beam no longer hit a grating. A
shift along $x$ is a shift along the stair-stepped direction. Shifts
along $x$ reduce $A_{\mathrm{eff}}$ for the same reason that the
layout is stair-stepped in the first place: Since photons come in at a
different angle, they need to be diffracted at a different distance
from the ML to hit the ML mirror at the Bragg peak.

For the rotation simulations, the origin of the rotation is the point
where the optical axes intersects the ``stair'' surface on which the
gratings are positioned. Of all the rotations, only rotations around
the $y$ direction (the dispersion axis) have limits tighter than a
degree or so, because rotation around $y$ changes the $z$ position of
the gratings, so the effect is similar to a translation in $z$.

\subsection{CAT gratings}
\begin{figure} [ht]
  \begin{center}
    \includegraphics[height=8cm]{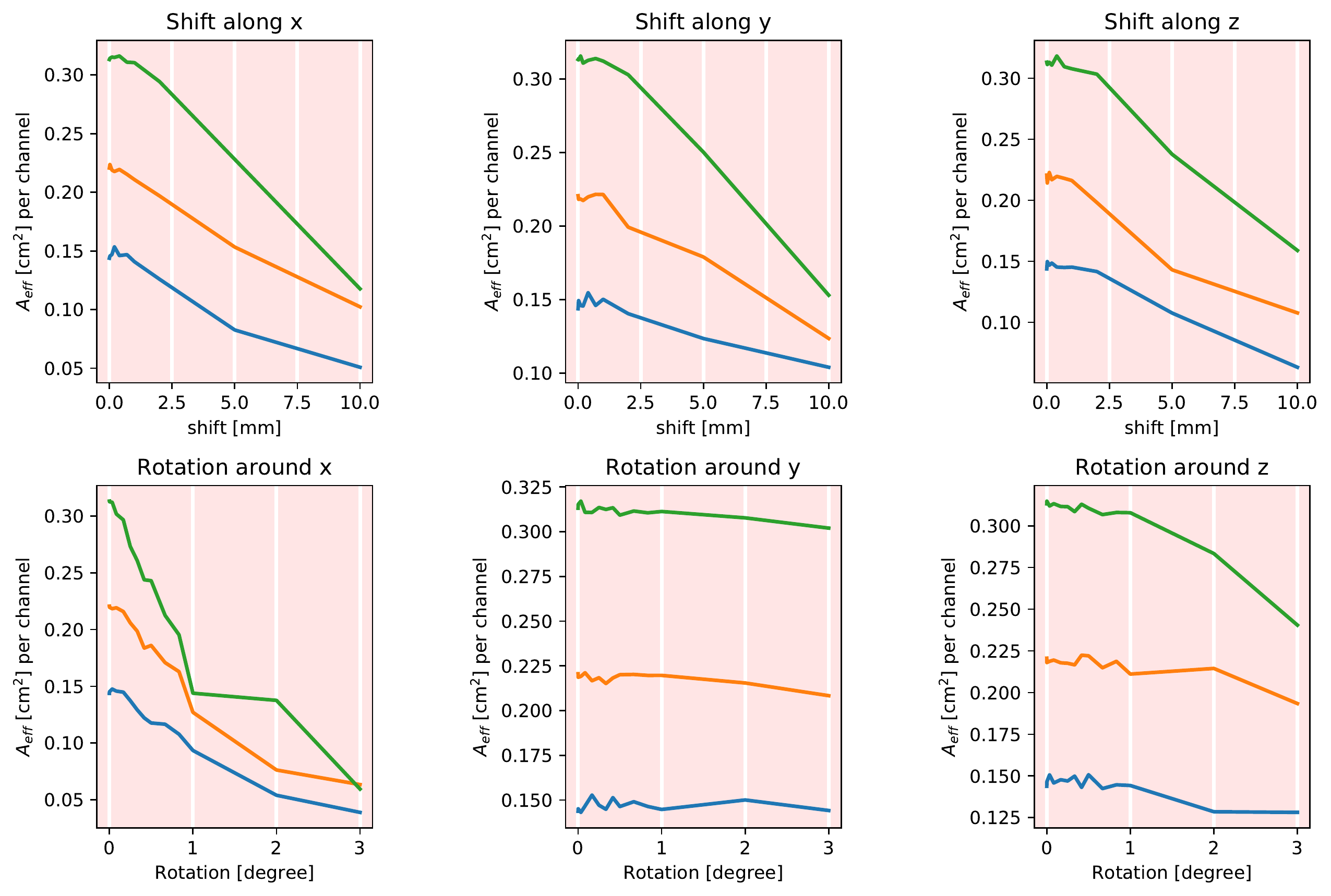}
  \end{center}
  \caption
      { \label{fig:CAT_individual}Change of effective area for CAT gratings misaligned with respect to each other.
}
\end{figure}

Figure~\ref{fig:CAT_individual} presents simulations where individual
CAT gratings are moved with respect to their nominal position on the
CAT gratings assembly. All translations allow $1\sigma$ errors of a few
mm, which is much larger than the size of the holder the gratings are
placed in. This is a trivial constraint. Similarly, only rotations
around $x$ (the short axis of the gratings) are tighter than 2
degrees. Rotations around $x$ makes the incoming photons hit the CAT
gratings at an angle different from the design blaze angle and reduce
the fraction of photons that are dispersed into the first order. Since
photons in other orders are not reflected from the ML mirror onto the
detector, this reduces the $A_{\mathrm{eff}}$ of the system. To keep the loss of $A_{\mathrm{eff}}$ below
10\%, the gratings need to be positioned within 10 arcmin of the
nominal rotation angle.

A change in the period of the gratings will also diffract photons to
the wrong locations, but the lithography process used to manufacture
the gratings gives a repeatability of the grating period that is
orders of magnitude better than the PiSoX requirement.

\subsection{ML mirror}
\begin{figure} [ht]
  \begin{center}
    \includegraphics[height=8cm]{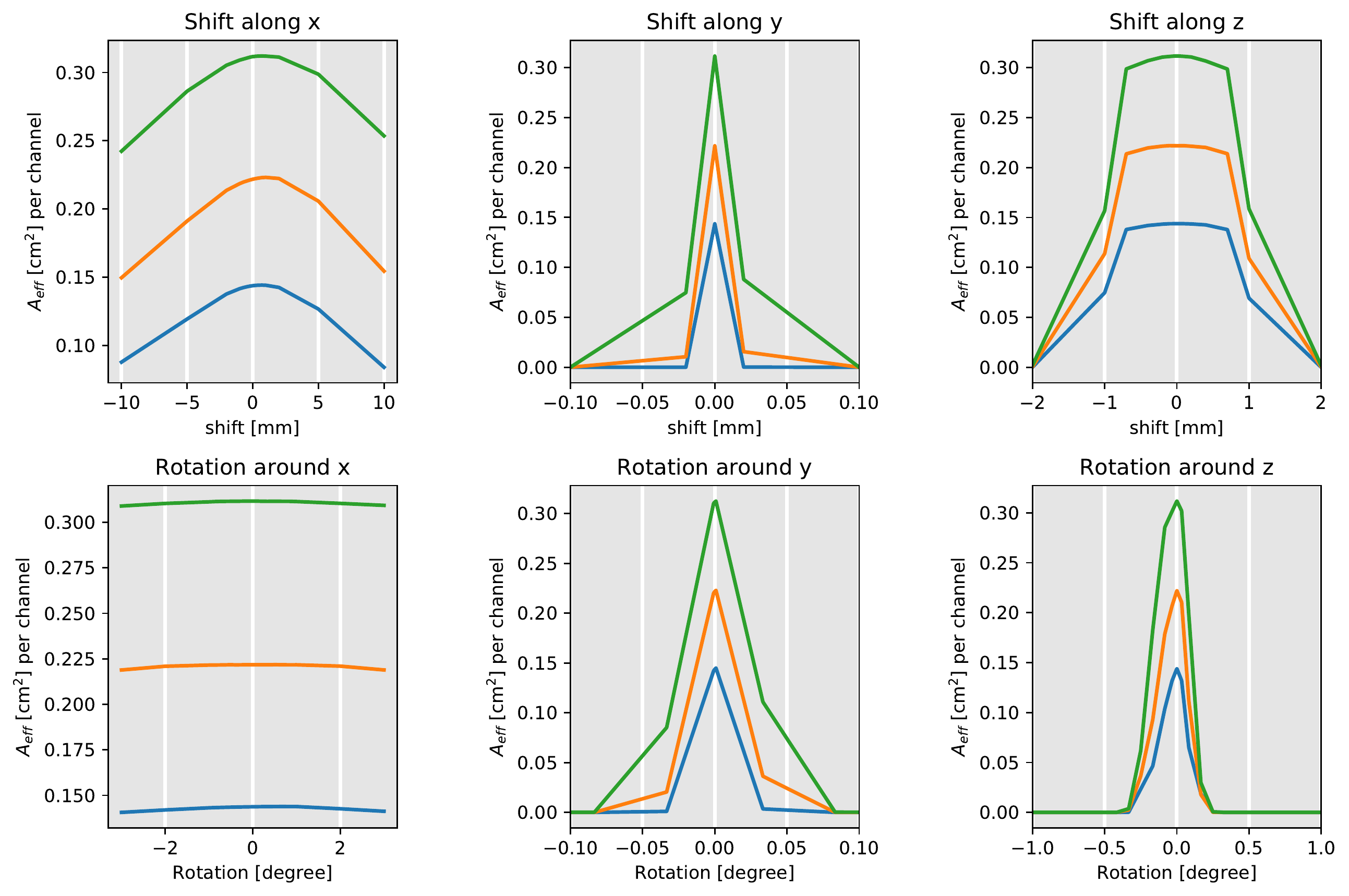}
  \end{center}
  \caption
      { \label{fig:LGML_global}Change of effective area in case of a misaligned ML mirror. 
}
\end{figure}

The ML mirror is the most critical part of the alignment because photons have
to hit the multilayer at the position of the Bragg peak. Because of
this, shifts along the $y$ direction (the direction in which the
multilayer is graded) are most sensitive
(Figure~\ref{fig:LGML_global}) and have to be aligned to better than
10 micron. However, this is for a source at nominal position. Moving
the source to a slightly off-axis position also changes where photons
interact with the ML mirror. Thus, in practice, this alignment does
not have to be performed to the 10~micron level for a single-channel
instrument. Instead sources can just be observed slightly off-axis to
compensate for any alignment error. This requires calibrating the
alignment in space by observing a source at different positions, until
the signal in the polarimetry channel is maximised. In this case, the
instrument can no longer rotate around the nominal axis to probe
different polarization angles. Instead, it has to rotate such that the
source is kept at the new position determined in the calibration. On
the other hand, tolerances for the other translations are a lot more
relaxed -- around a mm or so.

When there are two polarimetry channels,
there is generally an offset position
that will provide a match of the dispersed spectrum to the MLs
of both channels because
the effective area is so insensitive to cross-dispersion shifts.
For three detectors, a suitable shift may not be possible,
requiring that ground calibration properly align at least
two of the ML mirrors to each other.

Rotations around the long axis of the ML mirror ($x$ axis of the
coordinate system) have a very large tolerance of a few
degrees. Because the physical dimensions of the mirror are small, the
point of intersection with the mirror surface does not change much and
thus the photons still interact with the mirror very close to position
of the Bragg peak. On the other hand, rotations around the other two
axes move the mirror by a significant amount. That means that the
photons travel either too far or too little in $x$
direction, which causes them to miss the position of the Bragg peak
and consequently reduces $A_{\mathrm{eff}}$ significantly.

\begin{figure} [ht]
  \begin{center}
    \includegraphics[height=5cm]{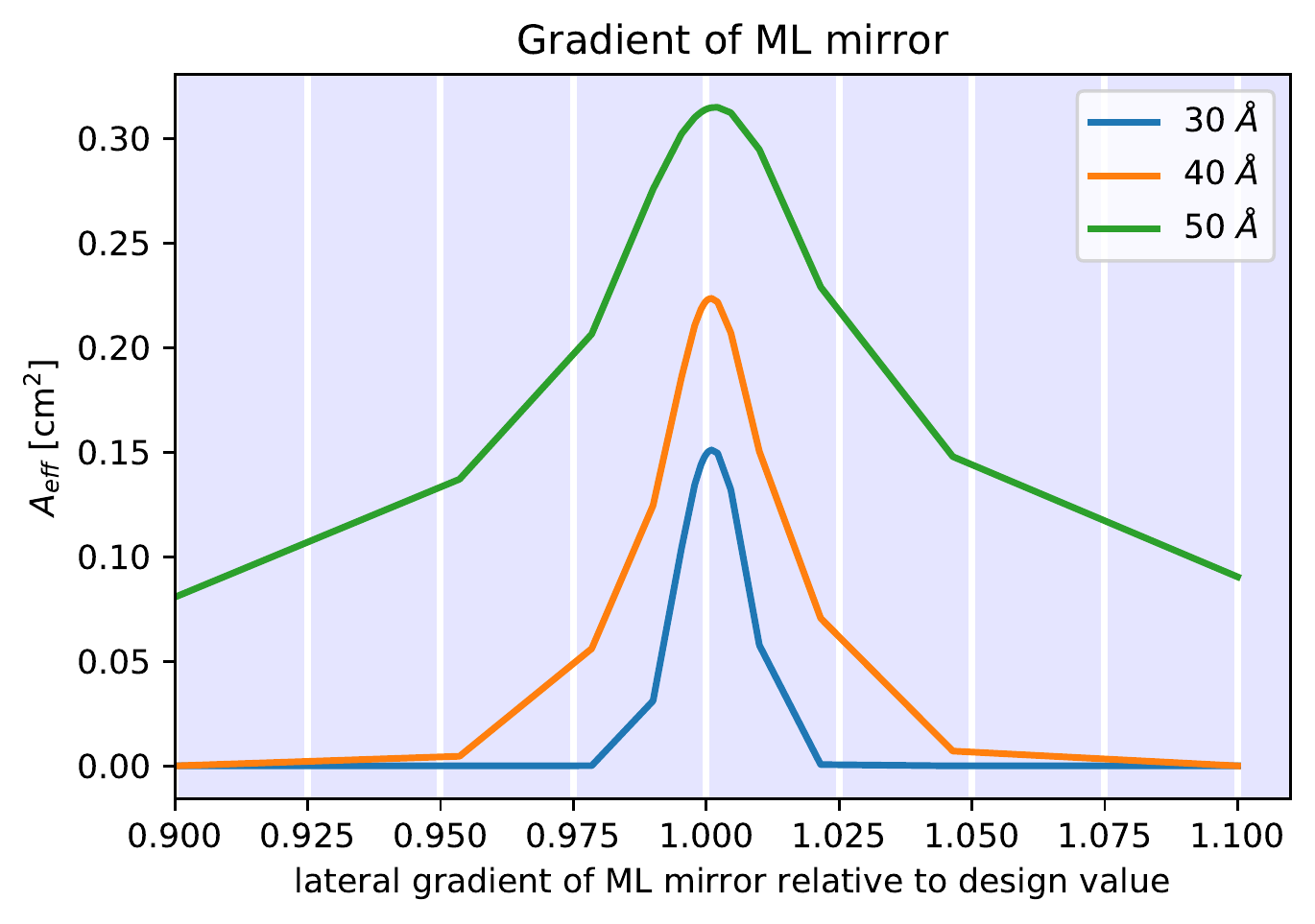}
  \end{center}
  \caption
      { \label{fig:LGML_gradient}Change of effective area for ML mirrors where the lateral grading differs from the design gradient.
}
\end{figure}

Another way for photons to miss the position of the Bragg peak is when
the lateral grading of the ML mirror does not match the gradient that
was assumed when placing the CAT
gratings. Figure~\ref{fig:LGML_gradient} shows that the gradient has
to be within about 1\% of the design value.

\subsection{Detectors}
The exact position of the detectors is not important as long as the
signal still hits the detectors. With increasing misalignment, the
spectral resolution of the polarimetry channel will degrade
slightly.
Background also increases
when the extraction region size needs to be increased, but again,
the background is negligible and the
effect of increasing the extraction region
size is negligible for any misalignment that can reasonably be
expected in the focal plane.

\section{SUMMARY}
\label{sect:summary}
One viable concept for a soft X-ray polarimeter is based on CAT
gratings and a multi-layer mirror. We show ray-traces for an
instrument designed for a small orbital mission with a focal length of
1.25~m and a single polarimetry channel. However, our results are
easily generalizable to additional channels, since each channel acts
essentially as its own instrument with separate CAT gratings, ML
mirrors, and detectors. We show the effective area and modulation factor
based on this design.

Gratings need to be bent to match the blaze angle in a converging
beam. We study different bending strategies and find that bending is
required, but the performance shows negligible sensitivity to the
exact value of the radius, so that only one type of grating holder
with the same radius for all gratings is sufficient, simplifying
design and handling. The requirement to place all gratings such that
diffracted rays hit the ML mirror at the position of the Bragg peak
limits the sizes of the CAT gratings. We find that a width of
10~mm optimizes performance.

Finally, we discuss alignment tolerances for all components. In most
degrees of freedom the requirements are so loose that they are well
below machining tolerances. However, the ML mirror
has to be positioned to 10~$\mu$m in the dispersion direction
relative to zeroth order, which can be calibrated and offset in flight.

\acknowledgments 
Support
for this work was provided in part through NASA grant NNX17AG43G and
Smithsonian Astrophysical Observatory (SAO) contract SV3-73016 to MIT
for support of the {\em Chandra} X-Ray Center (CXC), which is operated
by SAO for and on behalf of NASA under contract NAS8-03060.  The
simulations make use of Astropy, a community-developed core Python
package for Astronomy\cite{astropy1,astropy2}, numpy\cite{numpy}, and
IPython\cite{IPython}. Displays are done with mayavi\cite{mayavi} and
matplotlib\cite{matplotlib}.

\bibliography{report} 
\bibliographystyle{spiebib} 

\end{document}